\documentclass[prb,twocolumn,showpacs,preprintnumbers,amsmath,amssymb]{revtex4}
\usepackage{graphicx}
\usepackage{dcolumn}
\usepackage{bm}

\newcommand{\bes}{\begin{eqnarray}}
\newcommand{\ees}{\end{eqnarray}}
\newcommand{\be}{\begin{equation}}
\newcommand{\ee}{\end{equation}}

\begin{document}
\title{First-principles investigation of Ag-Cu alloy surfaces in an oxidizing environment}
\author{Simone Piccinin, Catherine Stampfl}
\affiliation {School of Physics, The University of Sydney, Sydney, New South Wales 2006, Australia}
\author{Matthias Scheffler}
\affiliation{Fritz-Haber-Institut der Max-Planck-Gesellschaft, Faradayweg 4-6, D-14195 Berlin, Germany}
\date{\today}

\begin{abstract}
In this paper we investigate by means of first-principles density functional theory calculations the (111) surface of the Ag-Cu alloy under varying conditions of pressure of the surrounding oxygen atmosphere and temperature. This alloy has been recently proposed as a catalyst with improved selectivity for ethylene epoxidation with respect to pure silver, the catalyst commonly used in industrial applications. Here we show that the presence of oxygen leads to copper segregation to the surface. Considering the surface free energy as a function of the surface composition, we construct the convex hull to investigate the stability of various surface structures. By including the dependence of the free surface energy on the oxygen chemical potential, we are able compute the phase diagram of the alloy as a function of temperature, pressure and surface composition. We find that, at temperature and pressure typically used in ethylene epoxidation, a number of structures can be present on the surface of the alloy, including clean Ag(111), thin layers of copper oxide and thick oxide-like structures. These results are consistent with, and help explain, recent experimental results. 
\end{abstract}

\maketitle

\section{Introduction}

Silver is the active element in heterogeneous catalysis for a number of industrially important chemical reactions including ethylene epoxidation (carried out at atmospheric pressure and temperatures $T$=500-600 K)~\cite{Santen87} and partial oxidation of methanol to formaldehyde (conducted at atmospheric pressure and $T$=800-900 K). The origin of the high catalytic activity of such a noble metal has been attributed to the thin oxide-like structures present on the lowest energy surface for realistic operating conditions.~\cite{LiPRL2003,MichaelidesJVCT05,SchloglPRB2003} In agreement with the Sabatier principle (in which a good catalyst readily dissociates adparticles but does not bind the fragments too strongly), it has been argued that while clean Ag(111) binds the adsorbates too weakly, the thin oxides provide O-Ag bonds of intermediate strength. A number of energetically very similar oxide-like structures have been identified for this system~\cite{MichaelidesJVCT05}, suggesting a scenario in which the surface might comprise of not simply a single low energy structure, but could well be dynamically evolving in time, fluctuating between different structures. This picture also emerges from recent studies on CO oxidation at RuO$_2$(110)~\cite{ReuterPRL04} and Pd(100)~\cite{RogalPRL07}: In both cases the highest catalytic activity is reached in regions of the phase diagram corresponding to boundaries between different stable surface oxide structures. Experiments conducted under semi-realistic catalytic conditions for CO oxidation at Pt(110) and Pd(100)~\cite{FrenkelCAT05} have also shown how, under steady-state catalysis, the thin oxides present at the surface continuously evolve with time. These recent results have therefore stressed how in catalysis, a non equilibrium process, the catalyst surface cannot be viewed as a static object but rather as a ``living'' system that, at the atomistic level, continuously evolves due to various processes such as adsorption, desorption, association, dissociation and diffusion. 
 
Having a detailed knowledge of the catalyst surface structures under operating conditions is therefore a crucial first step toward understanding the full catalytic process at the atomistic level. Conventional surface science investigations are conducted in ultrahigh vacuum (UHV) and at room temperature or below, while industrial applications of heterogeneous catalysis usually require pressure of the order of atmospheres and temperatures often higher than room temperature. The information extracted from such surface science experiments can not always be extrapolated to realistic conditions, since structures that can exist at high temperature and pressure might not be seen in UHV conditions and viceversa. Bridging this difference in ambient conditions (usually referred to as the pressure and temperature gap) is one of the major goals in current surface science research. To this end, first-principles simulations that combine accurate electronic structure methods with equilibrium thermodynamics have been successfully employed to predict the stability of surface structures of several systems under high temperature and high pressure conditions.~\cite{ReuterPRL03} Moreover, such methods have been included in a multiscale approach that employs statistical mechanics methods to simulate the full steady-state catalysis cycle. In some cases the agreement with experiments can be quantitative, allowing an unprecedented level of insight into the concerted actions of a dynamical process such as heterogeneous catalysis.~\cite{ReuterPRL04}        

With regard to the silver catalyst, the mechanism of ethylene epoxidation has been studied in two recent works~\cite{LinicJACS2002,LinicJACS2003} using a combination of experimental and theoretical techniques. It has been proposed that both the selective (producing ethylene oxide) and unselective (producing the acetaldehyde intermediate and finally leading to total oxidation) reaction pathways have a common intermediate, a surface oxametallacycle. This intermediate can react to form either ethylene oxide or acetaldehyde with similar activation barriers. The same authors also reported, on the basis of both first-principles calculations~\cite{LinicJC2004} and experiments~\cite{JankJC2005} that if an Ag-Cu alloy, rather than pure Ag, is used as a catalyst, the selectivity toward ethylene oxide will be improved.

The use of bimetallic catalysts like Ag-Cu has been the focus of much work in the field of heterogeneous catalysis,~\cite{SinfeltBOOK} since the catalytic activity and selectivity of a metal can be modified substantially by alloying with another metal. For example, geometrical and electronic effects obtained by varying the alloy composition may play a crucial role in determining the properties of the catalyst.~\cite{Liu01} One appealing aspect of bimetallic alloys is the possibility of rationally designing the catalytic properties of the material by changing the alloy elements and composition. To this end, first-principles calculations have been shown to be a potentially useful tool for screening among pools of possible alloys and for extracting trends and therefore gaining insights into the functioning of the alloy.~\cite{Greeley04, Greeley06}. In view of the above discussion, though, the effects of high pressure and temperature and the role of the dynamical evolution of the catalyst must be carefully accounted for, in order for the theoretical simulations to have predictive power.  

Through {\it ex-situ} X-ray photoelectron spectroscopy (XPS) measurements of Ag-Cu catalysts, it has been shown that the copper surface content is much higher than the overall copper content of the alloy,~\cite{LinicJC2004} therefore suggesting copper segregation to the surface. This lead Linic {\it et al.}~\cite{LinicJACS2002,LinicJACS2003} to theoretically model the surface of the Ag-Cu alloy assuming a perfect Ag(111) surface in which one out of four silver atoms is replaced by a copper atom. However, it is known that copper, at the temperature $T$ and pressure $p_{{\rm O}_2}$ used in such experiments ($T$=528~K, $p_{{\rm O}_2}$=0.1~atm), oxidizes to CuO,~\cite{Schmidt74} while at higher temperature or lower pressure Cu$_2$O is the stable oxide.~\cite{Schmidt74} Thus it is possible that more complex structures involving copper oxide can be present on the catalyst surface. 

The aim of this paper is to investigate, by means of first-principles electronic structure calculations, the surface structure of such alloys under varying oxygen pressure and temperature conditions. To do this we consider the alloy surface to be in thermodynamic equilibrium with an atmosphere of pure oxygen. The effect of the presence of other reactants such as ethylene is not investigated in this work. Under the conditions of steady-state catalysis it is therefore likely that the surface of the catalyst might be modified with respect to the stable structures found in an oxygen atmosphere. Hence our work can be thought of as a first step in gaining some insight in the possible structures of the alloy, an essential prerequisite for modeling the full catalytic process.  

The paper is organized as follows: In Sec.~\ref{Method} we briefly review the theoretical background relevant for this work, present the definitions of the quantities that will be used throughout the paper and discuss the approximations and assumptions that underpin our methodology. In Sec.~\ref{Results} we report our results and discuss their relevance. Section~\ref{Concl} summarizes the main findings.   
 
\section{Calculation Method}
\label{Method}

The density functional theory (DFT) calculations presented in this work are performed using the generalized gradient approximation (GGA) of Perdew-Burke-Ernzerhof (PBE)~\cite{PBE} for the exchange and correlation functional. We use ultrasoft pseudopotentials~\cite{USPP, USespresso} for the electron-ion interactions, including scalar relativistic effects. The Kohn-Sham wave functions are expanded in plane waves with an energy cutoff of 27 Ry (200 Ry for the charge density cutoff). To sample the Brillouin-zone we use the special-point technique,~\cite{Monkhorst-Pack} broadening the Fermi surface according to the Marzari-Vanderbilt cold-smearing technique,~\cite{MVcold} using a smearing parameter of 0.03 Ry. In the (1 $\times$ 1) surface unit cell, corresponding to the periodicity of clean Ag(111), a 12$\times$12$\times$1 \textbf{k}-point mesh is used: This amounts to 19 \textbf{k}-points in the irreducible part of the Brillouin-zone and gives adsorption energies for oxygen adsorption converged to within 16 meV\cite{convk} at full coverage. For larger unit cells, the \textbf{k}-point mesh is scaled accordingly. All the calculations are performed using the PWscf code contained in the Quantum-ESPRESSO package.~\cite{espresso}
For the calculations of chemisorbed structures we employ a 7 layer symmetric slab geometry, with adsorbates on both sides of the slab and fixing the position of the three central layers. To model oxide-like structures we place the thin copper layers on one side of a three layer Ag slab, fixing the bottom two layers~\cite{note_thickness}. The positions of the other atoms are relaxed until the forces are less than 0.001 Ry/au (0.025 eV/\AA).
A 12 \AA~vacuum layer is used, which is found to be sufficient to ensure negligible coupling between periodic replicas of the slab.~\cite{convV}
For the in-plane lattice spacing we use the calculated equilibrium bulk fcc lattice parameter of Ag, $a_0$ = 4.16~\AA. The experimental value for the lattice parameter is 4.09 \AA.~\cite{AshcroftMermin} 

We now define some quantities that will help the discussion presented in the following sections.
In the case of adsorption of oxygen on a clean Ag surface, the average binding energy per oxygen atom is defined as
\be
E^{{\rm O/Ag}}_b = -\frac{1}{N_{\rm O}}[E^{{\rm O/Ag}} - (E^{{\rm slab}} + N_{\rm O} E^{\rm O})]\quad,
\label{eq_eb}
\ee
where $N_{\rm O}$ is the number of oxygen atoms in the unit cell, $E^{{\rm O/Ag}}$, $E^{\rm slab}$ and $E^{\rm O}$ are the energies of the total system, the Ag slab and half the oxygen molecule (which has been computed through a spin-polarized DFT calculation), i.e. $E^{\rm O} = 1/2E^{\rm total}_{{\rm O}_2}$. Defined this way, a positive binding energy means that the adsorption of an oxygen atom is exothermic (i.e. stable) with respect to oxygen in molecular gas phase form. 
We point out here that the significant error ($\sim$ 0.54 eV) in the oxygen molecule binding energy introduced by the GGA-PBE approximation ($1/2E^{\rm O_2}_b = 3.10$ eV, while the experimental value is 2.56 eV~\cite{HuberBOOK}) would introduce a large error bar in the determination of the oxygen atom energy and therefore in the oxygen atom chemical potential. This error, however, will partially be compensated by the analogous GGA-PBE error in the description of oxygen chemisorbed on the metal surfaces.


To take into account the effects of temperature ($T$) and pressure ($p$) we employ ``{\it ab initio} atomistic thermodynamics'',~\cite{StampflCT2005, ReuterPRB2002, LiPRB2003} which allows the determination of the lowest energy structures as a function of the $T$ and $p$. The surface is considered to be in contact with an oxygen atmosphere that acts as a reservoir, therefore exchanging oxygen atoms with the surface without changing its temperature and pressure (i.e. its chemical potential).

The change in Gibbs free energy is calculated as  
\be
\Delta G(\mu_{\rm O}) = \frac{1}{A}(G^{\rm O/Ag} - G^{\rm{slab}}-\Delta N_{\rm{Ag}}\mu_{\rm{Ag}}-N_{\rm{O}}\mu_{\rm{O}}) \quad ,
\label{eq_gamma1}
\ee
where $G^{\rm O/Ag}$ is the free energy of the adsorbate/substrate structure, $G^{\rm slab}$ is the free energy of the clean Ag slab, $\Delta N_{\rm{Ag}}$ is the difference in the number of Ag atoms between the adsorption system and the clean Ag slab, $\mu_{\rm{Ag}}$ and $\mu_{\rm{O}}$ are the atom chemical potentials of Ag and O. The change in Gibbs free energy is normalized by the surface area $A$ to allow comparisons between structures with different unit cells; we will refer to this quantity as ``change in Gibbs surface free energy of adsorption'', or simply ``surface free energy''. The chemical potential of Ag is taken to be that of an Ag atom in bulk Ag, therefore assuming that the slab is in equilibrium with the bulk, that acts as the silver reservoir. The oxygen chemical potential depends on temperature and pressure according to~\cite{ReuterRuO2}
\be
\mu_{\rm O}(T,p) = \frac{1}{2}[ E^{\rm total}_{\rm O_2}(T, p^{\rm 0}) + \tilde\mu_{\rm O_2}(T,p^{\rm 0}) + k_BTln\left( \dfrac{p_{\rm O_2}}{p^{\rm 0}}\right)] \quad . 
\label{EqOxymu} 
\ee 
Here $p^{\rm 0}$ is the standard pressure and $\tilde\mu_{\rm{O}_2}(T,p^0)$ is the chemical potential at the standard pressure, which can obtained either from thermochemical tables~\cite{thermotables} (the choice made in this study) or directly computed.

The vibrational contributions to the free energy, which in principle should be accounted for, have been shown to be sufficiently small so as not to play an important role for the O/Ag system.~\cite{LiPRB2003, ReuterPRB2002} When comparing systems with different stoichiometry, on the other hand, vibrational contributions might play a non-negligible role. In this work however we will neglect such effects. 
To show that this approximation is valid, we have follow the procedure
described in Ref.~\cite{ReuterRuO2} to estimate the vibrational contribution
to the surface free energy, using the Einstein model and approximating the
phonon density of states by just one characteristic frequency for oxygen.
We find that for two of the most stable structures (``p2'' and
``CuO(1L)'', see Sec.~\ref{Results}) the oxygen characteristic frequencies
are 62 and 69 meV respectively, giving rise to differences in free surface
energy of less than 10 meV/\AA$^2$  in the range of temperatures of interest
(0-1000 K).
Hence, within this model, the only term that depends on $T$ and $p$ is the oxygen chemical potential. As a consequence, the free energies $G^{\rm O/Ag}$ and $G^{\rm slab}$ are identified as the total energies $E^{\rm O/Ag}$ and $E^{\rm slab}$. 

If we now consider the presence of copper impurities in the silver surface, we need to modify the definition of surface free energy. As we will show later in the paper, we will deal with structures in which an overlayer containing O, Ag and Cu is adsorbed on a clean Ag surface. If we consider Cu to be in equilibrium with a bulk Cu reservoir, the definition of surface free energy becomes:
\be
\Delta G(\mu_{\rm O}) = \frac{1}{A}(G^{\rm O/Cu/Ag} - G^{\rm slab}-\Delta N_{\rm{Ag}}\mu_{\rm{Ag}} - N_{\rm{Cu}}\mu_{\rm{Cu}} -N_{\rm{O}}\mu_{\rm{O}}) \quad ,
\label{eq_gamma2}
\ee
where $N_{\rm Cu}$ is the number of Cu atoms and $\mu_{\rm{Cu}}$ the copper chemical potential. 
Since in this work, as we will show in Sec.~\ref{Results}, we are
interested in identifying the structures belonging to the convex hull of the
free energy vs. copper content curve, which depends on the curvature of
such curve, the choice of the Cu chemical potential is arbitrary, since the
surface free energy depends linearly on it (see Eq.~\ref{eq_gamma2}).
We now define the average oxygen binding energy in the whole structure as
\bes
E^{{\rm O/Cu/Ag}}_b = && -\frac{1}{N_{\rm O}}[E^{{\rm O/Cu/Ag}} - (E^{{\rm slab}} + 
\Delta N_{\rm{Ag}}\mu_{\rm{Ag}} \nonumber \\
&& + N_{\rm{Cu}}\mu_{\rm{Cu}} + N_{\rm O} E^{\rm O})]\quad.
\label{eq_eb1}
\ees
Using the above-mentioned approximations, we can express the change in Gibbs surface free energy in terms of the average oxygen binding energy:
\bes
\Delta G(\Delta\mu_{\rm O}) = &&-\frac{1}{A}(N_{\rm O} E_b^{\rm O/Cu/Ag} + N_{\rm{O}}\Delta\mu_{\rm{O}}) \quad ,
\label{eq_gamma3}
\ees
where the oxygen chemical potential is now measured with respect to half an isolated O$_2$ molecule: $\Delta \mu_{\rm O} = \mu_{\rm O} - \frac{1}{2}E^{\rm total}_{\rm O_2}$.

In writing the change in free energy as in Eq.~(\ref{eq_gamma3}) we assume that the configurational entropy contribution to the free energy is negligible. To show that this is indeed the case, we compute the mixing entropy~\cite{DucastelleBook}
\be
S^{\rm mix} = -k_{\rm B}[x_s^{\alpha}ln(x_s^{\alpha}) + (1-x_s^{\alpha})ln(1-x_s^{\alpha})]\quad , 
\ee
where $x_s^{\alpha}$ is the concentration of the disordered phase $\alpha$ and $k_{\rm B}$ is the Boltzmann constant. The quantity $-TS^{\rm mix}$ around temperatures of interest for catalysis ($T$=600~K) has a maximum for $x_s^{\alpha}$=0.5, where it takes a value of 5 meV/\AA$^2$. As we will show later, such contribution is negligible compared to the energy scale of the surface free energies of the structures considered in this work. 
\section{Results}
\label{Results}

\subsection{Copper segregation to the surface}

Copper and silver are known to be almost completely immiscible in the bulk due to their large size mismatch (13\%): The maximum solubility limit is 8.2 mass \% Cu at the eutectic temperature of 779 $^\circ$C, and rapidly decreases at lower temperatures, with a value of 0.7 mass \% Cu at 400 $^\circ$C.~\cite{HansenBOOK} Accordingly, a DFT calculation at 0 K shows that the alloy mixing energy, defined as
\be
\sigma^{\rm mix} = (E^{\rm total} - N_{\rm Ag}\mu_{\rm Ag} - N_{\rm Cu}\mu_{\rm Cu})/(N_{\rm Ag}+N_{\rm Cu}), 
\ee
where $E^{\rm total}$ is the total energy of the unit cell of the alloy structure, is positive for all the concentrations considered here (between 1.4 \% and 6.2 \%). The results are shown in Table~\ref{TabAlloy}. Extrapolating the results at zero Cu concentration gives a positive mixing energy, therefore inhibiting alloying at 0 K at all concentrations. Here we relax all the atomic positions and keep the simulation cell fixed and commensurate to the computed equilibrium Ag lattice constant. 
 One would therefore expect Ag and Cu to phase segregate at low temperatures, where entropic effects are negligible.

\begin{table}[htb]
\begin{tabular}{ccccc}
\hline
\hline
cell  & $\%$Cu & $N_{\rm{Cu}}$ & $N_{\rm tot}$ & $\sigma^{\rm mix}$(meV/atom) \\
\hline
2$\times$2$\times$1 & 6.2 & 1 & 16 & 18.4 \\
2$\times$2$\times$3 & 4.2 & 2 & 48 & 12.6 \\
2$\times$2$\times$2 & 3.1 & 1 & 32 &  9.5 \\
2$\times$2$\times$3 & 2.1 & 1 & 48 &  6.5 \\
2$\times$3$\times$3 & 1.4 & 1 & 72 &  4.4 \\
\hline
\hline
\end{tabular}
\caption{Alloy mixing energy $\sigma^{\rm mix}$ as a function of the concentration of Cu impurities (in percent) in bulk Ag. The simulation cells used are indicated in the first column, given in terms of the conventional unit cell vectors, which have been kept fixed at the equilibrium Ag lattice constant. $N_{\rm tot}$ is the total number of atoms and $N_{\rm Cu}$ is the number of copper atoms present in the simulation cell.}
\label{TabAlloy}
\end{table}

\begin{figure}[htb]
\centering
\includegraphics[width=70mm]{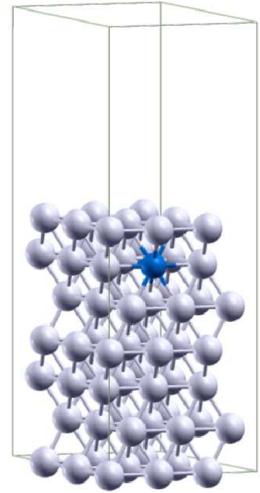}
\caption{\label{FigAg-Cu-3x3} (color online) The (3$\times$3) unit cell used for the study of segregation properties of Cu impurities in bulk Ag. The dark (blue) sphere represents a Cu atom.}
\end{figure}

If rather than bulk solid solutions we consider the presence of surfaces, we have to note that the surface energy of the most stable surfaces of Ag and Cu differ significantly:  The computed surface energy of Ag(111), 
$\gamma^{\rm Ag(111)}= 0.047$ eV/\AA$^2$ (exp.~\cite{BoerBOOK} $0.078$ eV/\AA$^2$), is much smaller than that of Cu(111), $\gamma^{\rm Cu(111)}= 0.076$ eV/\AA$^2$ (exp.~\cite{Lindgren84} $0.114$ eV/\AA$^2$). 

This suggests that, when copper impurities are introduced in silver, it is unlikely for copper to be exposed on the surface. To verify this, we compute the total energies of Ag slabs in which one substitutional copper atom is positioned in different layers, and all the atomic coordinates are fully relaxed, with the in-plane dimensions of the simulation cell fixed to the Ag lattice constant. We find that both bulk and surface positions are unfavored compared to the position directly below the first layer. These results are summarized in Table~\ref{TabSegr}. We can see that by increasing the size of the surface cell the qualitative picture is unchanged. In Fig.~\ref{FigAg-Cu-3x3} we show the unit cell used in the case of the (3$\times$3) cell. 

\begin{table}[htb]
\begin{tabular}{c|rrr}
\hline
\hline
Cu @ $N$-th Ag-layer & \multicolumn{3}{c}{$\Delta E$(meV)} \\
         & (1$\times$1) & (2$\times$2) & (3$\times$3) \\
\hline
1 &     56 &    93 &    76 \\
2 & $-$200 & $-$68 & $-$74 \\
3 & $-$103 & $-$38 & $-$39 \\
4 &      0 &     0 &     0 \\
\hline
\hline
\end{tabular}
\caption{Difference in total energy between the reference structure (where Cu occupies an Ag site in the center of the 7 layer slab, i.e. in layer 4) and with the Cu impurity positioned in other layers. The Ag-layer in which Cu sits is indicated in the first column. The (1$\times$1), (2$\times$2) and (3$\times$3) surface unit cells have been considered.}
\label{TabSegr}
\end{table}

\begin{figure}[htb]
\centering
\includegraphics[width=70mm, angle=270]{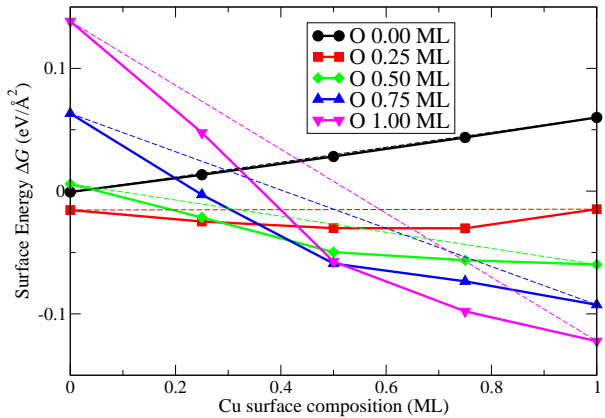}
\caption{\label{FigAg-Cusurfphasediag} (color online) Change in Gibbs surface free energy at $T = 0$ K (cf. Eq.~\ref{eq_gamma3}) as a function of Cu concentration in the first layer of Ag(111). The oxygen coverages considered are indicated in the legend in monolayers (ML). The energy zero corresponds the surface energy of the clean Ag(111) surface. The straight dashed lines are obtained by joining the points corresponding to pure Ag and pure Cu in the first layer.}
\end{figure}

 
\begin{figure*}[htb]
\begin{minipage}[c]{0.4\textwidth}
\centering
\includegraphics[width=60mm, angle=0]{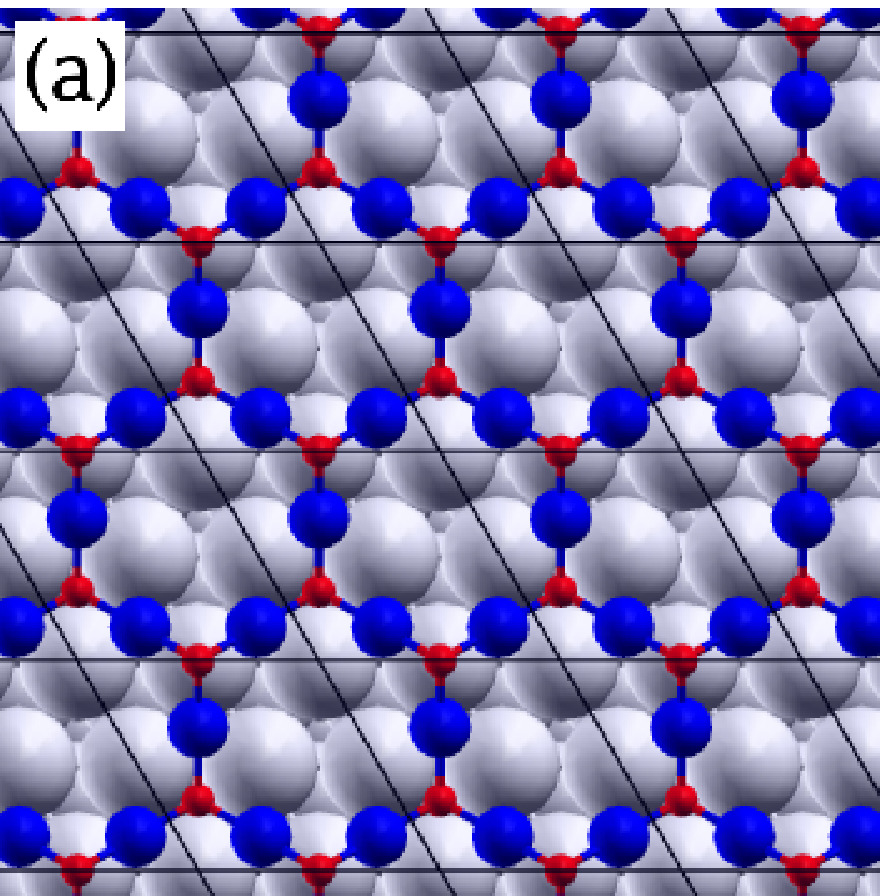}
\end{minipage}
\begin{minipage}[c]{0.4\textwidth}
\centering
\includegraphics[width=60mm, angle=0]{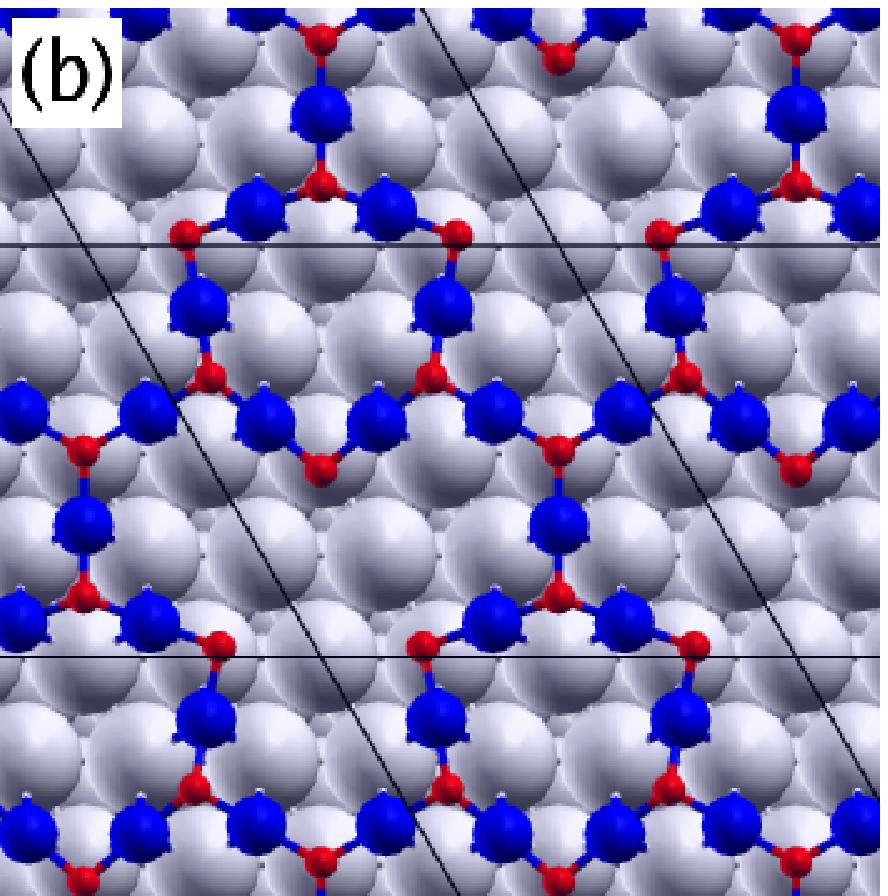}
\end{minipage}
\vskip 0.5cm
\begin{minipage}[c]{0.4\textwidth}
\centering
\includegraphics[width=60mm, angle=0]{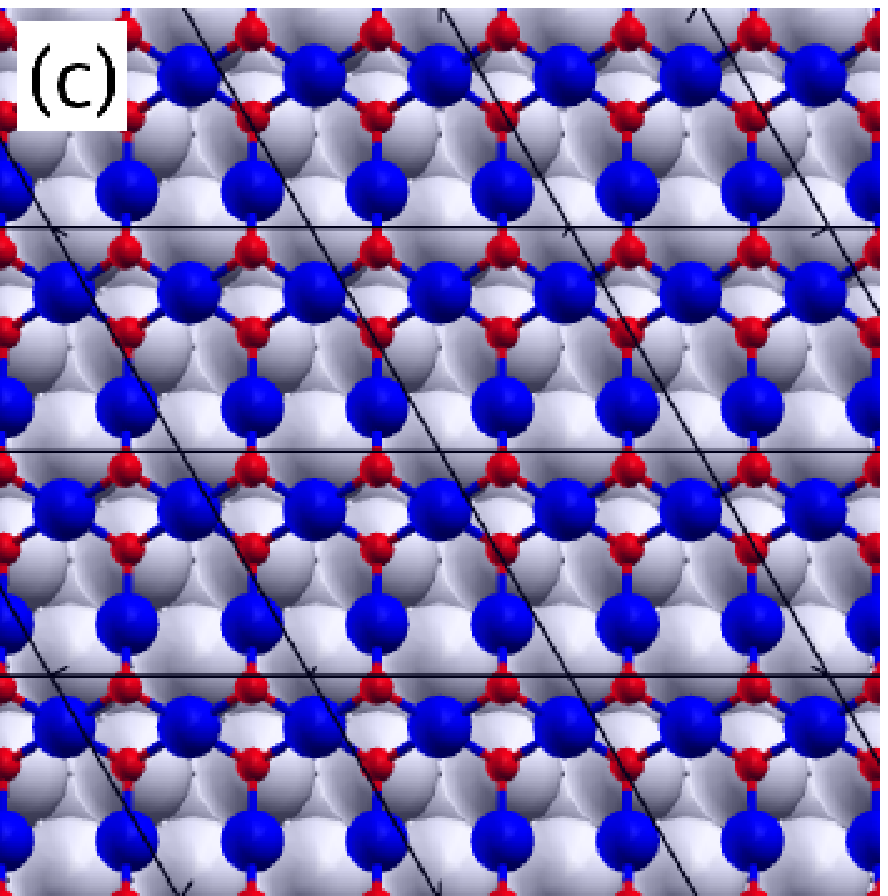}
\end{minipage}
\begin{minipage}[c]{0.4\textwidth}
\centering
\includegraphics[width=60mm, angle=0]{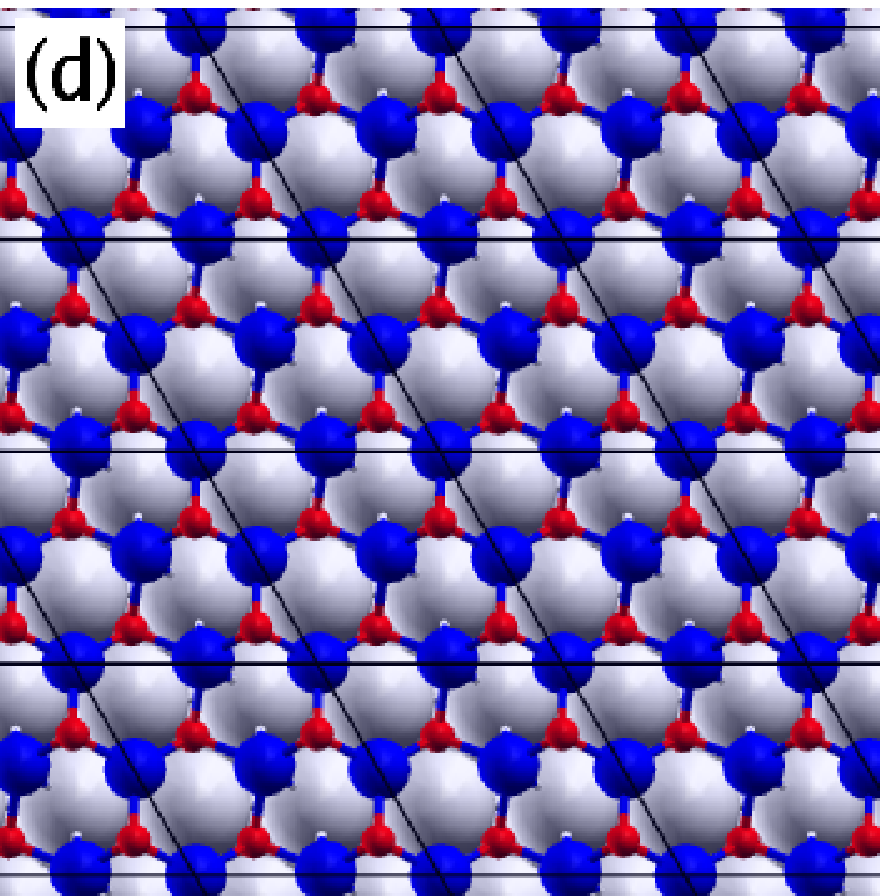}
\end{minipage}
\caption{\label{FigStruct} (color online) Top view of four surface structures considered: (a) ``$p$2'', (b) ``$p$4-OCu$_3$'', (c) ``CuO(1L)'' and (d)``O1ML/Cu1ML'' (see text). The grey spheres represent the underlying Ag(111) substrate. Copper atoms are shown as large dark (blue) circles, and oxygen atoms are the small dark (red) circles. The black lines represent the surface unit cells.}
\end{figure*} 

Accordingly, we also find that increasing the content of Cu in the first layer of the slab leads to an increase of the surface energy. This is shown in Fig.~\ref{FigAg-Cusurfphasediag} as the curve with black dots. The positive slope of the curve indicates that it is unfavorable for Cu impurities to migrate to the surface with respect to staying under the surface or in the bulk.~\cite{NorskovSurfAll} The fact that the curve is very slightly convex (i.e. the black dots lie below the straight line joining the two ends at composition 0 and 1), on the other hand, indicates that alloying in the first layer is slightly favorable with respect to phase separation into two structures that have on the first layer pure Ag and pure Cu. We note that on the scale of Fig.~\ref{FigAg-Cusurfphasediag} this is difficult to see, but is evident when plotted on a smaller scale.
This behavior (slightly convex curve) is what has been found in the theoretical work by Christensen {\it et al.}~\cite{NorskovSurfAll} for the Ag-Cu surface alloy on the Cu(100) surface, employing the linear muffin-tin orbitals (LMTO) method in the tight-binding representation, using the atomic-sphere approximation (ASA) and the coherent-potential approximation (CPA) and the LDA approximation for the exchange and correlation.~\cite{NorskovSurfAll} 

The small tendency to alloy and the large difference in surface energy between Ag(111) and Cu(111) have dramatic structural effects as has been recently shown from ultra high vacuum (UHV) experiments in which Cu thin-film growth on Ag(111) was studied.~\cite{MaurelSS2005,BocquetPRB2005} It was found that at room temperature, upon deposition of Cu on Ag(111), Cu forms islands that are encapsulated by one monolayer of Ag, in agreement with our DFT results presented in this work (cf. Table~\ref{TabSegr}).      

When oxygen is introduced into the picture, the situation changes completely: 
We find that the presence of oxygen chemisorbed on the alloy surface has the remarkable effect of reversing the slope of the (black dots) curve for the surface energy versus Cu surface composition, shown in Fig.~\ref{FigAg-Cusurfphasediag}. For these calculations we consider an Ag-Cu alloy in the first layer of the Ag slab, with oxygen adsorbed in the fcc site, i.e. in the most favorable adsorption site. Out of the four possible fcc sites in the (2$\times$2) cell, we always select the ones next to the copper atoms present on the first layer for the oxygen adsorption. Here we modelled the surface with a 7 layer slab with alloy structures created on both sides of the slab. By fully relaxing the structure, we calculate the change in surface free energy at $T = 0$ K as a function of the Cu content in the first layer and the oxygen coverage. The results are shown in Fig.~\ref{FigAg-Cusurfphasediag}. We can see that the presence of at least a quarter of a monolayer (ML) of oxygen induces Cu to segregate to the surface, i.e. changes the slope of the curve from positive (for an oxygen coverage relative to the underlying Ag(111) lattice $\theta_{\rm O} < 0.25$ ML) to negative (for $\theta_{\rm O} \geqslant 0.25$ ML). The driving mechanism here is the strong affinity between oxygen and copper, which more than compensates the unfavorable surface energy of Cu with respect to Ag: The adsorption energy (cf. Eq.(\ref{eq_eb})) of oxygen at 0.25 ML coverage on Ag(111) and Cu(111) is 0.38 eV/atom and 1.57 eV/atom, respectively. We also see that in the absence of copper (i.e. the left end of the graph in Fig.~\ref{FigAg-Cusurfphasediag}), the most stable structure is the one with $\theta_{\rm O} = 0.25$ ML. At higher oxygen coverages the oxygen-oxygen repulsion overcomes the formation of O-Ag bonds, leading to an increase in surface energy with the oxygen coverage. At the opposite end of the graph, where a full monolayer of copper is present in the first layer of the slab, the surface energy decreases with the oxygen coverage: This is due to the contribution to the surface energy of the formation of strong O-Cu bonds.    

\subsection{Surface alloy and surface oxide-like structures}

Having seen that oxygen induces copper to segregate to the surface, we now consider various surface structures with different contents of copper and oxygen in the first layer. These structures, formed on top of a pure silver slab, are periodic, and the largest surface unit cell is the (4$\times$4). As we will see in the next section, by changing the oxygen chemical potential according to Eq.~(\ref{EqOxymu}) we investigate the stability of these surfaces by evaluating the surface free energy at various temperatures and pressures. 

We consider three types of structures: ({\it i}) chemisorbed oxygen on the Ag-Cu alloy in the first layer of the Ag slab (as described in the previous section), ({\it ii}) structures derived from copper(I) oxide Cu$_2$O, whose structure can be visualized as trilayers of O-Cu-O piled up on top of each other and ({\it iii}) structures derived from copper(II) oxide CuO. 

The first set of structures are the ones considered in Fig.~\ref{FigAg-Cusurfphasediag}, and we label them O$x$ML/Cu$y$ML, where $x$ and $y$ are the content of O and Cu, expressed in monolayers with respect to the underlying Ag(111) surface. Some of these structures, in particular the ones with high Cu content, involve large atomic rearrangements that are due to the large Cu-Cu distances induced by the underlying Ag(111) lattice. 
As an example, we consider two structures, one at low oxygen coverage (O0.25ML/Cu1ML) and one at high oxygen coverage (O1ML/Cu1ML). In the first case we find that oxygen adsorbs at a very reduced height in the fcc hollow site of the (111) surface (0.43 \AA~compared to 1.22 \AA~for the O/Cu(111) system at the same coverage). In the second case, half the oxygen atoms penetrate below the copper layer, leading to a trilayer-like structure with a (2$\times$1) periodicity. In this structure half the oxygen atoms are positioned 0.95 \AA~ above the average plane of Cu atoms and half 0.86 \AA~below it. The atomic geometry is illustrated in Fig.~\ref{FigStruct}(d).

For the second set of Cu$_2$O-like structures, we use the label ``$p$2'' or ``$p$4'' depending on whether the periodicity of the structure is (2$\times$2) or (4$\times$4) with respect to the clean Ag surface. The atomic geometry of the ``$p$2'' structure is shown in Fig.~\ref{FigStruct}(a). Here each Cu is linearly bonded with two O and each O is bonded to three Cu, in a ring-like structure similar to the ones proposed for thin oxide-like layers on Ag(111)\cite{MichaelidesJVCT05} and Cu(111)\cite{SoonPRB2006}. 
The ``$p$4-OCu$_3$'' is a $p$4 structure in which an OCu$_3$ unit has been removed (see Fig.~\ref{FigStruct}(b)). In ``$p$4+Cu'' and ``$p$4+O'' a Cu atom and an O atom, respectively, has been added in an fcc adsorption site. ``Cu$_6$'' is analogous to the structure derived from the recently proposed ``Ag$_6$'' structure for the (4$\times$4)-O/Ag(111) system.~\cite{SchnadtPRL2006, SchmidPRL2006} We note that constraining an O-Cu-O trilayer of Cu$_2$O(111) to be commensurate with the $p$2 (or equivalently a $p$4) periodicity of Ag(111) results in a compression (7\%) of the trilayer with respect to its equilibrium lattice constant in bulk copper oxide. We also consider thicker films of Cu$_2$O (up to 5 atomic layers) in order to extrapolate the behavior of bulk Cu$_2$O.  
As we will show in the next section, the two most relevant thin oxide-like structures are ``$p$2'' and ``$p$4-OCu$_3$''. For these Cu$_2$O-like configurations, the average binding energy per oxygen atom is 1.41 eV and 1.37 eV, respectively, and the Cu-O bond length is between 1.84 and 1.85 \AA~in both cases. For comparison, in bulk Cu$_2$O the computed formation energy per oxygen atom is 1.26 eV (experimental value 1.75 eV~\cite{CRC}) and the Cu-O bond length is 1.88 \AA. 

For the third set of structures, we consider thin layers of CuO-like structures in a (2$\times$2) cell. Forcing the oxide layer to match the (2$\times$2) lattice of the underlying Ag(111) surface leads to a compression of the Cu-O bond length of about 3\% with respect to the bulk CuO value. 
For the one layer CuO structure (labelled ``CuO(1L)'', shown in Fig.~\ref{FigStruct}(c)) we find that the Cu-O bond length in the square planar pattern is 1.91 \AA~and the binding energy per oxygen atom is 1.16 eV. 
We consider a number of possible defective CuO-like thin layer structures (not shown here), but we find them not to be stable. 
We then consider thicker CuO-like films (up to 5 atomic layers) in order to extrapolate the behavior of bulk CuO. 
In this case we must bear in mind that bulk CuO is poorly described with DFT-PBE: CuO is a strongly correlated antiferromagnetic semiconductor, with a monoclinic structure ($a$=4.65 \AA, $b$=3.41 \AA, $c$=5.11 \AA, $\beta$=99.5$^\circ$), where Cu is linearly bonded to 2 O atoms and O is bound to 2 Cu.~\cite{KimJACS03} DFT-PBE, on the other hand, predicts CuO to be a metal with an almost orthorthorombic structure ($a$=4.34 \AA, $b$=4.01 \AA, $c$=5.22 \AA, $\beta$=92.2$^\circ$), in which each O is tetrahedrally coordinated to 4 Cu and each Cu is bonded in a square planar geometry to 4 O. 
The predictions for thick films of CuO must therefore be regarded as qualitative. The computed formation energy per oxygen atom in bulk CuO is 1.23 eV (experimental value 1.63 eV~\cite{CRC}) and the Cu-O bond length is about 1.97 \AA~(experimental value 1.67 \AA). 
Going beyond the DFT-PBE description is a major task. Calculations for CuO and
Cu$_2$O using GGA+U plus G$_0$W$_0$~\cite{JiangTBP} as well as other
methodology~\cite{HuReutPRL2007} are in progress.

In addition to the structures considered in this work, it is likely that other structures involving thin oxide-like layers with similar surface energies exist for this system. Recent works on O/Ag(111)\cite{MichaelidesJVCT05,SchnadtPRL2006, SchmidPRL2006} and O/Pd(111)~\cite{KlikovitsPRB2007} have shown that for these systems a multitude of structures with similar energetics can be found.   

\begin{figure}[tb]
\centering
\includegraphics[width=70mm, angle=270]{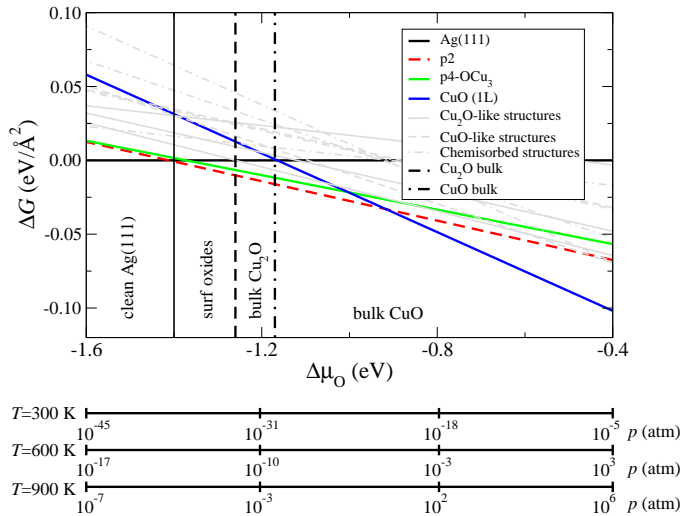}
\caption{\label{FigAg-Cuthermodiag} (color online) Change in Gibbs surface free energy as a function of the change in oxygen chemical potential $\Delta \mu_{\rm O}$. The chemical potential of oxygen is measured with respect to half the total energy of the free O$_2$ molecule. The vertical dashed lines separate the regions of stability of the clean Ag(111) surface, the surface oxides and bulk oxides Cu$_2$O and CuO.}
\end{figure}

\subsection{Thermodynamic diagram of the O/Cu/Ag(111) system}

In Fig.~\ref{FigAg-Cuthermodiag} we report the change in Gibbs surface free energy of the structures considered as a function of the change in oxygen chemical potential. We also show at the bottom of the plot, for three values of temperature (300, 600 and 900 K), the value of pressure corresponding to the chemical potential shown in the abscissa of the plot.

As we can see from Eq.~(\ref{eq_gamma3}), the slope of the lines in Fig.~\ref{FigAg-Cuthermodiag} is proportional to the oxygen coverage, i.e. the higher the oxygen content the steeper the line. The vertical dashed line at $\Delta \mu_{\rm O} = -1.26$ eV represents the change in chemical potential above which Cu oxidizes to Cu$_2$O, which, as one can show, corresponds to the computed heat of formation of Cu$_2$O. The vertical dashed line at $\Delta \mu_{\rm O} = -1.23$ eV, on the other hand, represents the change in chemical potential above which Cu oxidizes to CuO. We can therefore see that in the range of values of the oxygen chemical potential considered ($-1.6 < \Delta\mu_{\rm O} < -0.4$ eV) the thermodynamically stable structures are: Pure Ag for $\Delta\mu_{\rm O} < -1.43$ eV, ``$p$2'' and ``$p$4-OCu$_3$'' (almost degenerate) between  $-1.43 <\Delta \mu_{\rm O} < -1.26$ eV, bulk copper(I) oxide Cu$_2$O for $-1.26 <\Delta \mu_{\rm O} < -1.23$ eV and bulk copper(II) oxide CuO for $\Delta \mu_{\rm O} > -1.23$ eV. It is interesting to note that none of the alloyed chemisorbed structures (the structures labelled as O$x$ML/Cu$y$ML) are thermodynamically stable in this range of chemical potential, while there is a small region in which two-dimensional surface oxides are stable. This situation is similar to what has been found for the O/Cu(111) system,~\cite{SoonPRB2006} where it was argued that copper oxidation does not proceed via ordered chemisorbed structures, at variance with other transition metal structures. We also find that none of the O/Ag structures, containing no Cu (not shown in Fig.~\ref{FigAg-Cuthermodiag}) are thermodynamically stable in the range considered here. This is not unexpected, given the weaker strength of the Ag-O bond with respect to the Cu-O one.

\subsection{Modeling systems of known Cu surface content}

Having identified, among those considered here, the thermodynamically most stable ordered structures on an infinite Ag(111) surface, we now consider the situation in which we have a well defined Cu surface content. For a finite system (e.g. an Ag nanoparticle) of known dimensions and well defined Cu content, if we assume that all the Cu present in a spherical nanoparticle segregates to the surface in a site of the Ag fcc lattice, given the size of nanoparticle and the total Cu content, we can estimate the Cu surface content. For example in a 50 nm spherical nanoparticle with a 1~\% Cu content (typical values of nanoparticle size and copper content of Ag-Cu alloys used in experiments in which the effect of Cu impurities was investigated~\cite{JankJC2005}) the (average) surface Cu content is 0.35 ML, where 1 ML corresponds to the first layer of the nanoparticle being pure copper. Experiments carried out on Ag-Cu alloys with 0.1-1$\%$ Cu content at $\sim$ 500 K have shown that in an oxidizing environment the Cu surface content is in the range 0.1-0.75 ML.~\cite{JankJC2005}         

Given the fact that Ag(111) is the lowest energy surface for silver,~\cite{Vitos84} the typical understanding is that a large portion of the nanoparticle surface will consist of (111) planes. This is not necessarily true at high temperature, where the particle can be quite round, especially in the case of noble metals, and in the presence of oxide-like structures on the surface, that can alter the relative surface energies compared to the case of pure Ag at zero temperature. We will however restrict ourselves to this particular surface, and use the results shown in Fig.~\ref{FigAg-Cuthermodiag} to predict which structures will be present on the surface as a function of the copper content and the oxygen chemical potential. We also exploit the vast literature available for the O/Ag system~\cite{MichaelidesJVCT05, SchnadtPRL2006} to include the most stable structures for the system in the absence of copper. 

\begin{figure}[tb]
\centering
\includegraphics[width=70mm, angle=270]{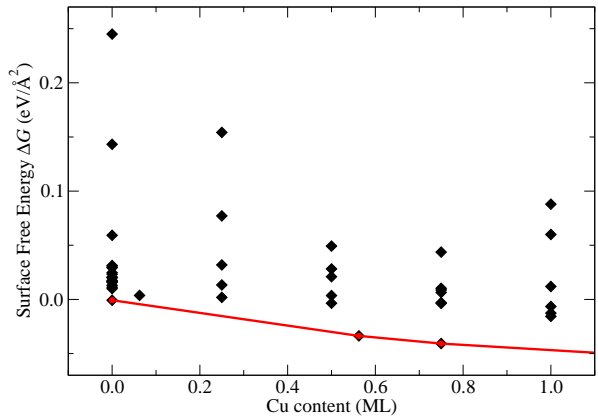}
\caption{\label{FigAg-Hull-0.8} (color online) Construction of the convex hull for the case of $\Delta\mu_{\rm O} = -0.80$ eV. The energy zero corresponds to the clean Ag(111) surface.}
\end{figure}

As an example, we show explicitly the case for $\Delta\mu_{\rm O} = -0.80$ eV, corresponding to an oxygen pressure of $10^{-3}$~atm at the temperature of 600 K. Figure~\ref{FigAg-Hull-0.8} shows the change in surface free energy at $\Delta\mu_{\rm O} = -0.80$ eV of all the structures we have considered (a total of 55 structures, including the O/Ag ones) as a function of the Cu surface content. By constructing the convex hull, i.e. the curve obtained by joining those structures that are stable with respect to linear combinations of structures at other compositions that would yield the same total composition, we identify the structures (indicated by the red dots in Fig.~\ref{FigAg-Hull-0.8}) that are stable against phase separation into any two other structures. The structures belonging to the convex hull, for this particular value of chemical potential, are ``Ag'' (clean Ag(111) surface), ``$p$4-Cu$_3$O'', ``$p$2'' and ``CuO(b)'' (bulk CuO). Although in Fig.~\ref{FigAg-Hull-0.8} we show only the portion of the diagram up to a Cu content of 1 ML, we have computed structures including up to five CuO layers in order to be able to extrapolate the behavior of bulk CuO. 
  
\begin{figure}[tb]
\centering
\includegraphics[width=70mm, angle=270]{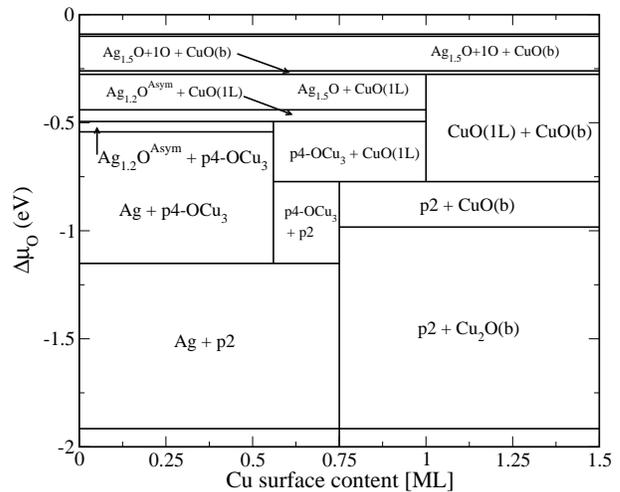}
\caption{\label{FigAg-structdiag} Surface phase diagram showing the structures belonging to the convex hull as a function of the Cu surface content and the change in oxygen chemical potential.}
\end{figure}

For a specific Cu content, e.g. 0.25 ML, the convex hull plot helps to predict that a mixture of pure Ag and ``$p$4-Cu$_3$O'' (rather than a single ordered structure with 0.25 ML Cu content) will be present on the Ag(111) surface at $\Delta\mu_{\rm O} = -0.80$~eV, in a ratio given by the lever rule, i.e. by the mass conservation law. By repeating this scheme for all the values of the oxygen chemical potential in the range considered in this work, we build a phase diagram as a function of the oxygen chemical potential and the copper surface content. This is shown in Fig.~\ref{FigAg-structdiag}. We find that the convex hull, for high Cu content, always includes the structure with the largest number of CuO layers (or Cu$_2$O layers at lower oxygen chemical potential), i.e. bulk oxide formation is favored beyond a certain Cu surface content. We therefore cut the plot in Fig.~\ref{FigAg-structdiag} at a Cu composition of 1.50 ML, where it is understood that the rightmost structure is ``bulk'' oxide. However, care must be taken in the meaning given to such ``bulk'' structure in the context of a finite system as the one we are modelling here. We interpret the numerical evidence of the presence of the bulk structure in the convex hull as a tendency for the formation of thick patches of either CuO or Cu$_2$O.

In the phase diagram shown in Fig.~\ref{FigAg-structdiag} we have explicitly written the two structures coexisting in a number of regions around the values of interest of oxygen chemical potential and copper content. 
The labels ``Ag$_{1.5}$O'' and ``Ag$_{1.2}$O$^{\rm Asym}$'' refer to O/Ag structures identified in Ref.~\onlinecite{MichaelidesJVCT05}.
If we focus on the region around the chemical potential of interest in typical industrial applications ($p$=1 atm, $T$=600 K, corresponding to $\Delta \mu_{\rm O} \sim -0.61$~eV) and for contents of copper below half a monolayer, we predict patches of one layer oxidic structures (``$p$4-Cu$_3$O'') to coexist with the clean Ag surface. At higher values of oxygen chemical potential, on the other hand, O/Ag structures can be found in coexistence with the ``$p$4-Cu$_3$O'' structure. For higher copper contents, the ``CuO(1L)'' and the ``p2'' structures are present in the phase diagram above and below $\Delta \mu_{\rm O} \sim -0.75$~eV, respectively. Finally, for even higher copper contents, bulk CuO is predicted to form.

This picture is consistent with recent experiments performed on the Ag-Cu system under catalytic conditions.~\cite{SpirosUNP} In these experiments Ag-Cu nanopowders ($~\sim$ 100 nm in diameter and 2.5\% Cu) are used as the catalyst for ethylene epoxidation at $T$ = 520 K and $p$ = 0.5 mbar. Through a combination of {\it in-situ} XPS, and Near Edge X-ray Absorption Fine Structure (NEXAFS) measurements, thin layers of CuO are found be to present on the surface, while no signs on a surface alloy are detected. Areas of clean Ag are also present on the surface, in agreement with our theoretical calculations.        

These results show that the simple structure adopted in Ref.~\onlinecite{LinicJC2004} for the Ag-Cu surface alloy to model theoretically the ethylene epoxidation reaction is not stable and is significantly different from what we expect to be relevant under high pressure conditions. The model used in Ref.~\onlinecite{LinicJC2004} assumes a (2$\times$2) periodicity for the Ag(111) surface in which one atom out of four is replaced with copper, and oxygen is chemisorbed on it. Our results suggest on the other hand that a model that includes, depending on the Cu surface content, clean Ag(111), patches of oxide-like thin layers and thick layers of CuO is a more appropriate model for the Ag-Cu catalyst (111) surface.

\section{Conclusions}
\label{Concl}

In summary, through density functional theory calculations and employing {\it ab-initio} atomistic thermodynamics to include the effects of pressure and temperature, we have investigated the (111) surface of the Ag-Cu alloy. We have shown that, in the absence of oxygen, copper impurities in silver prefer to stay directly below the silver surface, rather than in the bulk or on the surface. The presence of oxygen, on the other hand, has the effect to induce copper to segregate to the surface due to the strength of the O-Cu bond relative to the O-Ag one. Both findings are in agreement with experimental evidence. We have investigated structures involving chemisorbed oxygen on the alloy surface, as well as oxide-like structures on the Ag(111) surface. Through the construction of the oxygen chemical potential dependent convex hull, we have been able to identify, as a function of the surface copper content, the combinations of structures that are most stable at that particular temperature and pressure. In the region of interest, our results suggest that, depending on the copper surface content, clean Ag(111) and thin copper oxide-like structures (``$p$4-Cu$_3$O'', ``$p$2'' and ``CuO(1L)'') can coexist. This model for the (111) surface of the Ag-Cu catalyst differs substantially from the structures used in earlier works to investigate the effect of copper impurities in silver on the mechanism of ethylene epoxidation. Our results suggest that to gain insight into the full catalytic cycle one should consider oxygen species such as those in the proposed structures, rather than just oxygen chemisorbed on the the clean alloy surface. We envisage that the theoretical methodology reported in the present paper will also be useful for characterizing other alloy catalysts under realistic conditions.    
\bibliography{main}

\begin{thebibliography}{50}
\bibitem[{con({\natexlab{a}})}]{convk}
\bibinfo{note}{The value of 16 meV is obtained comparing the adsorption
  energies using the 19$\times$19$\times$1 and 30$\times$30$\times$1 meshes.}
\bibitem[{not()}]{note_thickness}
\bibinfo{note}{We checked the effect of increasing the thickness of the slab.
  In the case of the chemisorbed structures, going from 7 to 11 layer slabs
  leads to a change of the surface energy of the order of 2 meV/\AA $^2$. In
  the oxide-like structures, going from 3 to 5 layer slabs amounts to a change
  of the surface energy of the order of 1 meV/\AA $^2$.}
\end{thebibliography}

\end{document}